\documentclass[12pt,eqsecnum,preprint]{aastex}

\usepackage{amsmath}
\usepackage{graphicx}
\usepackage{tensor}

\def\pd#1#2{\frac{\partial #1}{\partial #2}}
\def\od#1#2{\frac{d #1}{d #2}}

\def\ahalf{{\frac{1}{2}}}

\def\prn#1{{\left(#1\right)}}
\def\brk#1{{\left[#1\right]}}
\def\brc#1{{\left\{#1\right\}}}

\begin{document}

\title{Nonaxisymmetric Neutral Modes in Relativistic Disks}

\author{Mike J. Cai}
\affil{Institute of Astronomy and Astrophysics, Academia Sinica \\
              P.O. Box 23-141, Taipei 106, Taiwan, R.O.C.}
\email{mike@asiaa.sinica.edu.tw}
\author{Frank H. Shu}
\affil{Department of Physics, National Tsing-Hua University \\
              Hsinchu 300, Taiwan, Republic of China}
\email{shu@mx.nthu.edu.tw}

\begin{abstract}
We perform a linear stability analysis of the axisymmetric,
relativistic, self-similar isothermal disk against non-axisymmetric
perturbations.  Two sets of neutral modes are discovered.  The first
set corresponds to marginally unstable perturbations driven by
gravitational radiation, and the other signals the onset of
bifurcation to non-axisymmetric equilibrium solutions to the Einstein
equations.
\end{abstract}
\section{Introduction}

\citet{CS:2002} put forward the hypothesis that relativistic disks
may ultimately play as important a role in astrophysics as their
spherical counterparts. There is evidence that supermassive
blackholes in quasars were formed during the early stages of
galaxy formation, where the major contributor to the mass-energy
density was gas rather than stars. The dynamics of gas is highly
dissipative, so it is easier for the gas to lose significant
amounts of energy to reach the desired compactness for general
relativistic effects to be important.  More problematic is how to
get rid of excess angular momentum if it is present initially.

The pioneering work of \citet{Bardeen:1971} on uniformly rotating
disks showed that with some assumed angular momentum loss, a Kerr
blackhole may result in the collapse of such disks. However, as
\cite{Mestel:1963} pointed out in a Newtonian context, there are
astrophysical reasons to think that a disk specified by constant
linear rotation velocity $v$ is more realistic than one which has
constant angular velocity.  Through bar formation and spiral
density waves \citep[for a review, e.g., see][]{Shu:2000}, such
differentially rotating disks possess natural mechanisms for the
outward transport of angular momentum and the inward transport of
mass that would promote, in the relativistic regime, blackhole
formation at the center.

The first step in a systematic theoretical study of this
possibility is to construct fully relativistic, self-similar,
rotating, flattened solutions.  \citet{LP} performed such a
pioneering analysis, but only in the cold limit. To suppress
well-known axisymmetric instabilities that would fragment the disk
into rings, it is necessary to include partial support from
isothermal pressure. If the pressure is exerted isotropically in
three directions, the result \citep{CS:2003} is the relativistic
generalization of the singular isothermal toroids (SITs) found by
\citet{Toomre:1982} and \cite{Hayashi:1982}.  A useful
approximation when such states become sufficiently flattened by
rotation is to ignore the thermal dispersive speed in the vertical
direction, while retaining it in the horizontal directions.  In
this approximation SITs become completely flattened, singular
isothermal disks (SIDs), whose equilibrium properties in the
relativistic regime were studied by \cite{CS:2002}. These
solutions are infinite in extent, possess infinite total mass, and
contain a naked singularity at the origin (a ``baby blackhole''
with vanishingly small mass).

The formal approximation of highly flattened configurations is
satisfied for SITs only when the Mach number $M \equiv
v/\sqrt{\gamma} \gg 1$.  Nevertheless, , even if $M \sim 1$,
\cite{CS:2003} found that the critical condition under which the
sequence of equilibria terminates as a function of $M$ is nearly
identical whether the pressure is exerted in three dimensions
(SITs) or two (SIDs).  The insensitivity of crucial properties of
the equilibria to the assumption of infinitesimal thickness will
hopefully carry over to the analysis of their stability.

On dimensional grounds, if the disk becomes gravitationally
unstable to overall gravitational collapse (a possibility if the
disk is sufficiently slowly rotating), the mass of the baby black
hole will grow linearly in time as a result of axisymmetric
collapse. In the analogous problem of the collapse of relativistic
singular isothermal sphere (SIS), \cite{CS:2004} have shown that
the growth of a black hole with finite mass introduces a
(spherically symmetric) horizon which covers up the singularity.
It is intriguing to ask whether such singularity in the case of a
collapsing, relativistic SID will remain naked if the requirement
of axial symmetry is relaxed. In order to answer this question,
one must first construct relativistic SIDs that are
non-axisymmetric states of equilibria.  One of the goals of the
present paper is to make a start on this problem, by finding the
points of non-axisymmetric bifurcation along the sequence of
axisymmetric SIDs, thereby generalizing the Newtonian work of
\citet{Syer:1996} and \citet{Galli:2001}.

A feature with no Newtonian analog appears with the completion of
the analysis: the appearance of a secular instability which
afflicts all relativistically rotating disks. This instability,
associated classically with the name Rossby (so the corresponding
perturbations are called R-modes), arises because general
relativity admits the radiation of angular momentum (and energy)
by gravitational waves \citep{Chandra:1970a, Chandra:1970b}.
Basically, if the rotation of the underlying axisymmetric state is
high enough, a counter-rotating disturbance appears corotating to
an inertial observer.  Such disturbances have negative angular
momentum density in the local rest frame of the disk. Since
gravitational radiation carries away positive angular momentum,
the amplitude of the R-mode perturbation grows in time.

The phenomenon renders, in some sense, all astrophysically
rotating systems potentially unstable to spin-down on a
gravitational-radiation time scale. Whether the R-mode instability
competes with the gravitational torques associated with barlike
deformations or spiral density waves remains a problem for future
study.  The self-similar models and techniques used in the present
paper are capable only of determining the criterion for the onset
of secular instabilities, and not their growth rates and evolution
into the nonlinear regime.  Thus, modifications are still
required for application to astrophysically
realistic circumstances, where the origin does not contain a
singularity from the start, and where spacetime at infinity is
flat.

In this paper, we restrict our study of the stability of
relativistic SIDs to non-axisymmetric perturbations with the same
scale-free character as the equilibrium state (i.e., with the same
power-law radial dependence). In the nomenclature of
\citet{Syer:1996}, we consider only {\it aligned} perturbations,
and no {\it spiral} disturbances. In section \ref{axisymmetric},
we review the basic properties of axisymmetric SIDs. In section
\ref{perturbed config}, we develop the mathematical formulation of
the stability analysis, including the metric and matter
perturbation.  In section \ref{Newtonian}, the equations are
solved in the Newtonian limit, and the result is compared to
\cite{Shu:2000}.  In section \ref{result}, the perturbation
equations are solved in the full relativistic context, and we offer
physical interpretation of the results.

\section{Review of Axisymmetric Disk Solution}\label{axisymmetric}
Start out with a self-similar axisymmetric metric
\begin{equation}
    ds^2 = -r^{2n} e^N dt^2 + r^{2} e^{2P-N} (d\phi - r^{n-1} e^{N-P}
Q dt)^2 + e^{Z-N} (dr^2 + r^2 d\theta^2),\label{bifur-metric}
\end{equation}
where $N$, $P$, $Q$, $Z$ are functions of $\theta$ and $n$ is a
constant measuring the strength of the gravitational field. For
numerical convenience, we have chosen the equatorial plane to be
at some polar angle $\theta_0$, which is determined as an
eigenvalue of the problem. The locally nonrotating observer (LNRO)
defines an orthonormal tetrad frame analogous to the inertial
frame:
\def\tet#1#2{\tensor{e}{_{(#1)}^{#2}}}
\newcommand{\et}{\tilde \varepsilon}
\begin{equation}
\begin{split}
    \tet{0}{\mu} &= (r^{-n} e^{-\ahalf N}, r^{-1}Q e^{\ahalf
N-P}, 0, 0)\\
    \tet{1}{\mu} &= (0, r^{-1}e^{\ahalf N-P}, 0, 0)\\
    \tet{2}{\mu} &= (0, 0, e^{\ahalf(N-Z)}, 0)\\
    \tet{3}{\mu} &= (0, 0, 0, r^{-1}e^{\ahalf(N-Z)}).
\end{split}\label{LNRO}
\end{equation}
We look for solutions to the Einstein field equations with a disk
matter source described by a constant linear rotation velocity and
a two-dimensional isotropic pressure. In the frame of a LNRO, the
stress-energy tensor is taken to be
\def\T#1#2{T_{(#1)(#2)}}
\begin{equation}
\begin{split}
    &\T{0}{0} = \frac{\varepsilon + p_\phi v^2}{1-v^2}, \quad \T{0}{1} =
-\frac{(\varepsilon+p_\phi)v}{1-v^2},\\
    &\T{1}{1} = \frac{p_\phi+ \varepsilon v^2}{1-v^2}, \quad \T{2}{2} =
    p_r,
\end{split}
\end{equation}
where $\varepsilon \propto \delta(\theta- \theta_0)$ and $p_\phi =
p_r = \gamma \varepsilon$.  Define
\begin{equation*}
    \Theta = (1+n)\theta, \qquad ' = \od{}{\Theta},
    \qquad \et = 8\pi \frac{\varepsilon}{1+n} r^2 e^{Z_0-N_0},
    \qquad \Delta = \delta(\Theta-\Theta_0).
\end{equation*}
After some algebra, part of the Einstein equations are cast into a
set of dynamic equations
\begin{equation}
\begin{split}
    &N'' =- N' P' - \frac{2n}{1+n} +Q^2 F^2
+ Q^2 \prn{\frac{1-n}{1+n}} ^2+
\et\brk{\frac{1+2\gamma+v^2}{1-v^2}}\Delta, \\
    &P'' = -{P'}^2-1+\et\gamma \Delta,\\
    &Q'' = -Q' P' + Q \brk{(1-Q^2) \prn{\frac{1-n}{1+n}}^2 + (N'-P')^2
    -Q^2 F^2}\\
    &\phantom{Q''=}-\et(1+\gamma) \frac{2v+Q+Qv^2}{1-v^2}\Delta,\\
    &Z''= - Z' P' + 2Q^2\prn{\frac{1-n}{1+n}}^2
- \frac{4n^2}{(1+n)^2} + Q^2 F^2+2\et\brk{\frac{v^2 +
\gamma}{1-v^2}} \Delta,\\
\end{split}\label{unperturbed-dyn}
\end{equation}
where $F = N'+\log Q' -P'$.  The rest of the field equations and
the equation of motion form a set of constraint equations
\begin{equation}
\begin{split}
    &Q(\Theta_0) v(1-n)(1+\gamma) + \gamma + v^2 - n (1+\gamma) =
    0, \\
    &Z'-\frac{2n}{1+n} N'- Q^2\frac{1-n}{1+n} F = 0,\\
    &Q^2\brc{F^2-\prn{\frac{1-n}{1+n}}^2} + 2
    \cot \Theta Z' - {N'}^2 + \frac{4n^2}{(1+n)^2} = 0.
\end{split}\label{unperturbed-constraint}
\end{equation}
Due to the contracted Bianchi identity, two of these equations are
redundant, which were used for a numerical consistency check. For
detailed discussion on the properties of these solutions, the
readers are invited to refer to \cite{CS:2002}.

\section{Perturbed Configuration}
\label{perturbed config} We will use the Eulerian description for
the non-axisymmetric modes.  The perturbation in the metric is
\begin{equation*}
    \delta g_{\mu\nu} = h_{\mu\nu}, \qquad
    \left |\frac{h_{\mu\nu}}{g_{\alpha\beta}} \right | \ll 1.
\end{equation*}
We define the change in the contravariant components of the metric
as
\begin{equation*}
    \delta g^{\mu\nu} = - h^{\mu\nu} = - h_{\alpha\beta} g^{\alpha
    \mu} g^{\beta\nu}
\end{equation*}
so that\footnote{It is unfortunate that we have two $\delta$
symbols here -- one denoting Eulerian change and the other
denoting the Kronecker-Delta function.  There should be little
confusion from context, however.}
\begin{equation*}
    (g^{\mu\nu} + \delta g^{\mu\nu}) (g_{\nu\rho} + \delta
    g_{\nu\rho}) = \tensor{\delta}{^\mu_\rho}+ O(h^2).
\end{equation*}
Notice that $h_{\mu\nu}$ is \textit{not} a tensor with respect to
the unperturbed metric, and its projection onto LNRO is not a
Lorentz scalar. Thus the directional derivatives of $h_{(a)(b)}$
will involve more than the usual Ricci rotation coefficients,
which destroys the simplicity of the tetrad formalism.  As a
result, we will compute in the coordinate frame whenever
derivatives are involved.  However, the tetrad frame defined by
\eqref{LNRO} does offer a clean separation of $r$ from the other
coordinates, so we shall project our results onto LNRO
\textit{after} the derivatives have been taken.

The change in the Ricci tensor reads \citep[see,
e.g.,][]{Wald:1984}
\begin{equation}
    \delta R_{\mu\nu}=\ahalf \prn{\tensor{h}{^\alpha_{\mu;\nu\alpha}}
    + \tensor{h}{^\alpha_{\nu;\mu\alpha}} -
    \tensor{h}{_{\mu\nu;}^\alpha_\alpha} -
    h_{;\mu\nu}},
\end{equation}
where the raising and lowering of indices are done with the
unperturbed metric.  Instead of computing the Einstein tensor, we
work out the trace-reversed stress-energy tensor.  Taking the
direct variation of the stress-energy tensor, we have
\begin{equation*}
    \delta(T_{\mu\nu} - \ahalf g_{\mu\nu} T) = \delta T_{\mu\nu} -
    \ahalf h_{\mu\nu} T - \ahalf g_{\mu\nu} \delta T^{\alpha\beta}
    g_{\alpha\beta} - \ahalf g_{\mu\nu}
    T^{\alpha\beta}h_{\alpha\beta}.
\end{equation*}
There is one subtlety in writing down this expression, due to the
non-tensorial nature of the metric variations $h_{\mu\nu}$.
Explicitly,
\begin{equation*}
    \delta T_{\mu\nu} = \delta T^{\alpha\beta}
    g_{\alpha\mu}g_{\beta\nu} + \tensor{T}{_\mu^\alpha} h_{\alpha
    \nu} + \tensor{T}{_\nu^\alpha}h_{\alpha\mu} \ne
    \delta T^{\alpha\beta}
    g_{\alpha\mu}g_{\beta\nu}.
\end{equation*}
To avoid such confusion, we shall adopt the convention that only
contravariant components $T^{\alpha\beta}$ are
varied\footnote{This is not entirely unfamiliar. Recall that in
the super-Hamiltonian formalism, the conjugate momentum to $x^\mu$
is $p_\mu$, which is what we vary, not $p^\mu$.}.  So the Einstein
field equation reads
\begin{equation}
\begin{split}
    &\tensor{h}{^\alpha_{\mu;\nu\alpha}}
    + \tensor{h}{^\alpha_{\nu;\mu\alpha}} -
    \tensor{h}{_{\mu\nu;}^\alpha_\alpha} -
    h_{;\mu\nu}\\
    = &8\pi (2\delta T^{\alpha\beta}
    g_{\alpha\mu}g_{\beta\nu} + 2\tensor{T}{_\mu^\alpha} h_{\alpha
    \nu} + 2\tensor{T}{_\nu^\alpha}h_{\alpha\mu}-h_{\mu\nu} T
    - g_{\mu\nu}\delta T^{\alpha\beta}g_{\alpha\beta} - g_{\mu\nu}
    T^{\alpha\beta}h_{\alpha\beta} ).
\end{split}\label{Einstein}
\end{equation}

\subsection{Gauge Choice}
Let's consider perturbations with time and angular dependence
$e^{im\phi - i\omega t}$.  On dimensional grounds, a scale-free
disk can not support modes with $\omega \ne 0$.  The limiting case
$\omega = 0$ signals the onset of bifurcation or marginal
stability of a particular mode.  The most general form of
$h_{(a)(b)}$ may be written as
\begin{equation*}
    h_{(a)(b)} =
    \begin{pmatrix}
    h_{00}&h_{01}&h_{02}&h_{03}\\
    h_{01}&h_{11}&h_{12}&h_{13}\\
    h_{02}&h_{12}&h_{22}&h_{23}\\
    h_{03}&h_{13}&h_{23}&h_{33}
    \end{pmatrix}e^{im\phi},
\end{equation*}
where the $10$ $h$ entries are functions of $\theta$ only.
Geometrically, the metric coefficients are the inner products of
the basis vectors
\begin{equation*}
    g_{\mu\nu} = \pd{}{x^\mu} \cdot \pd{}{x^\nu}.
\end{equation*}
Since we are only considering polar perturbation, the system is
symmetric about the equator and the metric is invariant under the
diffeomorphism $\theta \rightarrow 2\theta_0-\theta$.  This
implies the boundary condition
\begin{equation}
    h_{t\theta} = h_{\phi\theta} = h_{r\theta} = 0 \Rightarrow  h_{\mu 3}
= 0 \text{ for } \mu \neq 3 \label{metric-bc}
\end{equation}
on the disk.

We proceed as follows.  Project the left-hand side of
\eqref{Einstein} onto the LNRO, and write the result as
$l_{(a)(b)}(1+n)^2e^{im\phi} e^{N-Z} /r^2$. Expand $l_{(a)(b)}$
and replace the second derivatives of zeroth-order metric
coefficients with the unperturbed Einstein equations
\eqref{unperturbed-dyn}.  This will introduce singular terms on
the disk.  Since we require the metric to be continuous across the
disk, all singular terms must balance for the first-order
equations in $l_{(a)(b)}$, which are $l_{(a)3}$ (the second-order
equations are acceptable since the first derivatives will in
general have a jump there). Miraculously, with the condition
\eqref{metric-bc}, all first-order equations are regular.

To proceed further, we need to choose a gauge. Consider an
infinitesimal coordinate transformation $x^\mu \rightarrow y^\mu=
x^\mu + \xi^\mu(x)$, where $\xi^\mu$ is of the same magnitude as
$h_{\mu\nu}$.  This induces a transformation on the metric in the
usual way,
\begin{equation*}
\begin{split}
    g_{\alpha\beta}(x) &= \pd{y^\mu}{x^\alpha} \pd{y^\nu}{x^\beta}
    g'_{\mu\nu}(x+\xi) = (\delta^\mu_\alpha +
    \tensor{\xi}{^\mu_{,\alpha}})(\delta^\nu_\beta +
    \tensor{\xi}{^\nu_{,\beta}})(g'_{\mu\nu} + \xi^\rho
    g'_{\mu\nu,\rho})\\
    &= g'_{\alpha\beta} + g'_{\mu\alpha}
    \xi^\mu_{,\beta} + g'_{\mu\beta} \xi^\mu_{,\alpha} +
    \xi^\rho g'_{\alpha\beta,\rho} = g'_{\alpha\beta} + (
    \pounds_{\boldsymbol{\xi}}\mathbf{g'})_{\alpha\beta} =
    g'_{\alpha\beta} + 2\xi_{(\alpha;\beta)}.
\end{split}
\end{equation*}
Thus, the coordinate freedom we have in general relativity
corresponds to the gauge freedom $h_{\mu\nu} \rightarrow
h_{\mu\nu} + 2 \xi_{(\mu;\nu)}$. As suggested by the boundary
condition on the disk, we shall promote \eqref{metric-bc} to a
gauge condition. There is one more degree of freedom which we will
fix here.  The total gauge thus reads
\begin{equation}
    h_{03}=h_{13}=h_{23} = 0, \qquad h_{11} = h_{33}.\label{gauge}
\end{equation}
The last condition resembles the Regge-Wheeler gauge in spherical
symmetry. With the gauge condition, we may write
\begin{equation}
\begin{split}
        h_{tt} &= r^{2n}e^N [a+Q^2b - 2Q d]
        e^{im\phi}, \qquad
        h_{t\phi} =r^{n+1}e^P (d- Qb)e^{im\phi},\\
        h_{tr} &= i r^n e^{Z/2} (f-Qj)e^{im\phi}, \qquad
        h_{\phi\phi} =r^2e^{2P-N}be^{im\phi} = e^{2P-Z} h_{\theta\theta},\\
        h_{\phi r} &=ire^{P+Z/2-N}je^{im\phi}, \qquad
        h_{rr} =e^{Z-N}ce^{im\phi},
\end{split}
\end{equation}
which corresponds to
\begin{equation}
h_{(a)(b)} = e^{im\phi}
\begin{pmatrix}
    a&d&if&0\\
    d&b&ij&0\\
    if&ij&c&0\\
    0&0&0&b
\end{pmatrix}.
\end{equation}
The left-hand side of \eqref{Einstein} now reads
\allowdisplaybreaks{
\begin{align*}
    l_{00}&=-a''-(\ahalf N'+P')a' + \ahalf N' c' + 2QF d' \\
    &+\brc{-Q^2\prn{\frac{1-n}{1+n}}^2-Q^2F^2-\et
    \Delta\frac{1+2\gamma+v^2}{1-v^2}+e^{Z-2P}\frac{m^2}{(1+n)^2}}a\\
    &+\brc{-2Q^2\prn{\frac{1-n}{1+n}}^2-Q^2F^2+\frac{2n}
    {1+n}+2e^{Z-2P}Q^2\frac{m^2}{(1+n)^2} - \frac{1+2\gamma+v^2}{1-v^2} \et
    \Delta}b\\
    &+\brc{Q^2\prn{\frac{1-n}{1+n}}^2 - \frac{2n}{1+n} + Q^2e^{Z-2P}
    \frac{m^2}{(1+n)^2}}c\\
    &+2\brc{Q\brk{\prn{\frac{1-n}{1+n}}^2 - e^{Z-2P}\frac{m^2}{(1+n)^2}+(N'-P')F}
    - \frac{2v(1+\gamma)}{1-v^2} \et \Delta}d\\
    &-e^{Z/2-P}\frac{2Qm(1+2n)}{(1+n)^2}f + e^{Z/2-P}
    \frac{2m(n+nQ^2-Q^2)}{(1+n)^2} j,\\
    l_{11}&=-b''+(P'-\ahalf N') a' - P' b' +(\ahalf N'-P') c'
    +2QF d'\\
    &+\brc{-e^{Z-2P}\frac{m^2}{(1+n)^2}-Q^2\brk{\prn{\frac{1-n}{1+n}}^2
    +F^2}}a\\
    &+\brc{e^{Z-2P}\frac{(1-Q^2)m^2}{(1+n)^2}-Q^2\brk{2\prn{\frac{1-n}{1+n}}^2
    +F^2}-\frac{2}{1+n}}b\\
    &+\brc{e^{Z-2P}\frac{m^2}{(1+n)^2}+Q^2\prn{\frac{1-n}{1+n}}^2
    +\frac{2}{1+n}}c\\
    &+2Q\brc{e^{Z-2P}\frac{m^2}{(1+n)^2}+\prn{\frac{1-n}{1+n}}^2
    +(N'-P')F}d\\
    &+2Qe^{Z/2-P} \frac{m}{(1+n)^2}f - 2 e^{Z/2-P}
    \frac{m}{(1+n)^2} [Q^2(1-n)+n+2]j,\\
    l_{22} &=-c''+\ahalf(Z'-N')a' + \ahalf (N'-Z'-2P') c' +
    Q^2\prn{\frac{1-n}{1+n}}^2 a\\
    &+\brc{2Q^2\prn{\frac{1-n}{1+n}}^2+\frac{2n(1-n)}{(1+n)^2}
    -\et \Delta} b -2Q\prn{\frac{1-n}{1+n}}^2 d\\
    &+ \brc{(1-Q^2)e^{Z-2P} \frac{m^2}{(1+n)^2}+
    \et\Delta - Q^2\prn{\frac{1-n}{1+n}}^2-
    \frac{2n(1-n)}{(1+n)^2}}c\\
    &+Qe^{Z/2-P} \frac{2mn}{(1+n)^2} f +e^{Z/2-P}
    \frac{2m}{(1+n)^2} (Q^2-1-nQ^2) j,\\
    l_{01} &=-d'' + \ahalf QF a' + Q F b' +\ahalf QF c' - P' d'\\
    &+Q\brc{e^{Z-2P} \frac{m^2}{(1+n)^2} - 2\frac{1-n}{(1+n)^2}}
    b + Q\brc{e^{Z-2P} \frac{m^2}{(1+n)^2} + 2\frac{1-n}{(1+n)^2}}
    c\\
    &+\brc{(1-Q^2)\prn{\frac{1-n}{1+n}}^2-\et\Delta \gamma +
    (N'-P')^2 - Q^2F^2}d\\
    &-e^{Z/2-P}\frac{2mn}{(1+n)^2} f + 2Qe^{Z/2-P}
    \frac{m(n-2)}{(1+n)^2}j,\\
    il_{02} &= f'' +P' f'- QF j'+\brc{Qe^{Z-2P}\frac{m^2}
    {(1+n)^2}+\prn{P'-\ahalf Z'}QF+2\frac{(1+\gamma)v}{1-v^2}
    \et \Delta}j\\
    &+e^{Z/2-P}\frac{m}{(1+n)^2}\brc{\ahalf Q(1-n) a + Q(1-n) b
    - \ahalf (n+3) Q c - (1-n) d}\\
    &+\brc{-e^{Z-2P}\frac{m^2}{(1+n)^2}-(N'-\ahalf Z')^2
    +\frac{2n}{(1+n)^2}+\ahalf Q^2F^2+ \frac{v^2+\gamma}{1-v^2}
    \et \Delta}f,\\
    il_{12} &= j'' - QF f' +P' j'+Q\brc{-e^{Z-2P}
    \frac{m^2}{(1+n)^2}-2\frac{1-n}{(1+n)^2}
    +(\ahalf Z' -N')F}f\\
    &+e^{Z/2-P}\frac{m}{(1+n)^2}\bigg\{(1-n)(1+\ahalf Q^2)a +
    2(1-n)Q^2b\\
    &- [1+n+\ahalf Q^2(1-n)]c - 3(1-n) Q d\bigg\}\\
    &+\bigg\{e^{Z-2P}Q^2\frac{m^2}{(1+n)^2} +Q^2\prn{\frac{1-n}{1+n}}^2
    +\frac{1+2n-n^2}{(1+n)^2}\\
    &-(P'-Z'/2)^2+\ahalf Q^2 F^2
    - \frac{1+\gamma v^2}{1-v^2}\et \Delta\bigg\}j.
\end{align*}
}

\subsection{Matter Content}
Recall that the disk is made of a two-dimensional perfect fluid.
Explicitly, if we choose the equation of state $p = \gamma
\varepsilon$, the unperturbed stress-energy tensor may be written
as
\begin{equation}
    T^{\mu\nu} = \varepsilon\brk{(1+\gamma) u^\mu u^\nu + \gamma
g^{\mu\nu}}, \text { for } \mu, \nu = t, \phi, r, \text { and }
T_{\lambda \theta} = 0.
\end{equation}
In the presence of a perturbation, we still need to impose the
condition that momentum flux and stress in the vertical direction
vanish.  Hence the first-order change in the stress-energy tensor
is only for the upper-left $3\times 3$ block:
\begin{equation*}
    \delta T^{\mu\nu} = \delta \varepsilon \brk{(1+\gamma) u^\mu u^\nu
+ \gamma g^{\mu\nu}} +\varepsilon \brk{(1+\gamma)(\delta u^\mu
u^\nu + u^\mu \delta u^\nu) - \gamma h^{\mu\nu}}.
\end{equation*}
As usual, the four-velocity is normalized,
\begin{equation*}
    (u^\mu + \delta u^\mu) (u^\nu +\delta u^\nu) (g_{\mu\nu} +
h_{\mu\nu}) = -1 \Rightarrow \delta u^\mu u_\mu = -\ahalf u^\mu
u^\nu h_{\mu\nu}.
\end{equation*}
Projecting onto the LNRO frame, we have
\begin{equation}
    \delta T_{(a)(b)} = \delta \varepsilon[(1+\gamma) u_{(a)} u_{(b)}
+ \gamma \eta_{(a)(b)}] + \varepsilon[(1+\gamma)(\delta u_{(a)}
u_{(b)} + u_{(a)} \delta u_{(b)}) - 2\gamma w_{(a)(b)}]
\end{equation}
and
\begin{equation*}
    \delta u^{(a)} u_{(a)} = - u^{(a)} u^{(b)} w_{(a)(b)},
\end{equation*}
where
\begin{equation*}
    u^{(a)} = (\frac{1}{\sqrt{1-v^2}},
\frac{v}{\sqrt{1-v^2}}, 0, 0), \qquad \delta u^{(a)} = (\frac{x
e^{im\phi}}{\sqrt{1-v^2}},
    \frac{y e^{im\phi}}{\sqrt{1-v^2}}, \frac{ize^{im\phi}}{\sqrt{1-v^2}}
    , 0).
\end{equation*}
With this parameterization, the normalization condition reads
\begin{equation}
   x=(y+d)v + \ahalf (a+v^2b).
\end{equation}
Next, we work out the equation of motion (EOM) for the perturbed
quantities.  Although it is not needed if we solve all six unknown
metric perturbations directly, the EOM provides a consistency
check for the algebraic mess that is notorious in general
relativity. Furthermore, as we will see later, for the
self-similar disk, the EOMs are all algebraic, which are much
easier to handle than the full Einstein equations. A direct
variation of $\tensor{T}{^{\mu\nu}_{;\nu}}=0$ reads
\begin{equation*}
    \tensor{\delta T}{^{\mu\nu}_{;\nu}} +
    \tensor{\delta \Gamma}{^\mu_{\rho\nu}}
    T^{\rho \nu} + \tensor{\delta\Gamma}{^\nu_{\rho\nu}} T^{\rho\mu}=
    0,
\end{equation*}
where
\begin{equation*}
    \tensor{\delta \Gamma}{^\mu_{\nu\rho}} = \ahalf g^{\mu\alpha}
    (h_{\alpha\nu;\rho} + h_{\alpha\rho;\nu} -
    h_{\rho\nu;\alpha}).
\end{equation*}
Just as in the unperturbed EOM, the $\mu=\theta$ component needs
to be satisfied identically.  Physically, this component gives the
fluid evolution in the $\hat \theta$ direction, which is trivial
in this case.  If we assume that the metric coefficients are even
about the equatorial plane, then the first derivatives are odd,
and vanish upon integrating through the plane.  The resulting
equation only contains $h_{\nu\theta}$ where $\nu\neq 3$.  This is
another reason why the full gauge has to satisfy the boundary
condition \eqref{metric-bc}. Next, we consider the density
perturbations. The zeroth-order density that is self-similar may
be written as
\begin{equation*}
    \varepsilon = \frac{A}{r^2} \delta(\theta-\theta_0),
\end{equation*}
where $A = \et e^{N_0-Z_0}/8\pi$ is some constant. Conventionally,
the equatorial plane is located at $\theta=\pi/2$ in spherical
polar coordinates. However, to reduce eigenvalues of the problem,
we rescaled $\theta$ so that the disk is located at $\theta_0$.
When the geometry is perturbed, this ``hidden'' eigenvalue will in
general change, hence we need to vary $\theta_0$ as well.  Thus,
we obtain
\begin{equation}
\begin{split}
    \delta \varepsilon &= \frac{\delta A e^{im\phi}}{r^2}
    \delta(\theta-\theta_0) + \frac{A}{r^2} [\delta (\theta
    -\theta_0 -\theta_1 e^{im\phi}) - \delta(\theta-\theta_0)]\\
    &=\frac{e^{im\phi}}{r^2} \brk{\delta A \delta(\theta-\theta_0) -
    A\delta'(\theta-\theta_0) \theta_1}.
\end{split}
\end{equation}
Projecting the right-hand side of \eqref{Einstein} onto the LNRO
and writing the result as $H_{(a)(b)}(1+n)^2 e^{im\phi}
e^{N-Z}/r^2$, the nonzero components are\footnote{Actually
$H_{33}$ is nonzero too, but it contains exactly the singular
terms from $l_{33}$.}
\begin{equation}
\begin{split}
    H_{00}&=\frac{1+2\gamma+v^2}{1-v^2} \zeta \Delta +
    \Big [-\frac{1+\gamma v^2+Qv(1+\gamma)}{1-v^2}a +
    \frac{v(2v-Q)(1+\gamma)}{1-v^2}b \\
    &+ \frac{(v-Q)(1+\gamma)
    -Qv^2(1+\gamma)}{1-v^2}d +4v\frac{1+\gamma}{1-v^2} y
    \Big ]\et\Delta,\\
    H_{01}&=-\frac{2v(1+\gamma)}{1-v^2} \zeta \Delta
    + \Big [-v\frac{(2+v^2)(1+\gamma)}{1-v^2}b
    - \frac{2(1+\gamma v^2) + \gamma + v^2}{1-v^2}d \\
    &- 2\frac{(1+\gamma)(1+v^2)}
    {1-v^2}y\Big ]\et \Delta,\\
    iH_{02} &=\et \Delta \brk{\frac{\gamma + \gamma v^2 + 2}{1-v^2} f
    + 2v\frac{1+\gamma}{1-v^2} j + 2\frac{1+\gamma}{1-v^2} z},\\
    H_{11}&= \frac{1+v^2+2\gamma v^2}{1-v^2} \zeta \Delta +
    \brk{\frac{\gamma+3v^2 + 2\gamma v^2}{1-v^2} b
    +3v\frac{1+\gamma}{1-v^2}d + 4v\frac{1+\gamma}{1-v^2} y
    }\et \Delta,\\
    iH_{12} &=-\et \Delta \brk{2v\frac{1+\gamma}{1-v^2} f
    + \frac{\gamma +\gamma v^2 + 2v^2}{1-v^2}j
    + 2v\frac{1+\gamma}{1-v^2}z},\\
    H_{22} &=\zeta \Delta + \gamma c \et \Delta,
\end{split}
\end{equation}
where
\begin{equation*}
    \zeta \Delta = 8\pi \delta \varepsilon \frac{r^2 e^{Z-N}}{1+n}
e^{-im\phi}.
\end{equation*}
If we use the definition of $\delta \varepsilon$, the above
expression simplifies to
\begin{equation*}
    \zeta \Delta = \et_1 \Delta - \et e^{N_0 - Z_0} e^{Z-N}
\Delta' \Theta_1 = \et_1 \Delta + \et (Z'_0 -N'_0) \Theta_1 \Delta
\Rightarrow \zeta = \et_1 + \et (Z'_0 -N'_0) \Theta_1,
\end{equation*}
where we integrated by parts on the last term.  In fact, the
explicit form of $\zeta$ is not required here since $\et_1$ and
$\Theta_1$ never appear independently in the equations.  This
observation suggests that $\Theta_1$ is a second order effect.

\subsection{Boundary Conditions}
>From symmetry, all the non-axisymmetric metric components should
vanish on the axis (where $\Theta=0$).  Thus we can impose the
conditions
\begin{equation}
    a=b=c=d=f=j=0.\label{BC-axis}
\end{equation}
On the disk, the delta functions give rise to a jump in the
derivatives of the perturbation functions using the second-order
equations. Integrating across the disk, we have
\begin{equation}
\begin{split}
    2a'&=\et\bigg\{\frac{v^2+2\gamma-\gamma
    v^2-Qv(1+\gamma)}{1-v^2}a+\frac{1+2\gamma+3v^2+2\gamma v^2
    -Qv(1+\gamma)}{1-v^2}b\\
    &+\frac{(1+\gamma)(5v-Q-Qv^2)}{1-v^2}d +\frac{4v(1+\gamma)}
    {1-v^2} y \bigg\} + \frac{1+2\gamma+v^2}{1-v^2}\zeta,\\
    2b'&=\et \brc{\frac{2\gamma v^2+3v^2+\gamma}{1-v^2} b
    +\frac{v(1+\gamma)}{1-v^2} (3d+4y) } + \frac{1+v^2+2\gamma
    v^2}{1-v^2} \zeta,\\
    2c'&=\et \brc{b-(1-\gamma)c} + \zeta,\\
    2d'&=\et \brc{-\frac{v(v^2+2)(1+\gamma)}{1-v^2} b-
    \frac{3\gamma v^2+2+v^2}{1-v^2} d - \frac{2(1+v^2)(1+\gamma)}
    {1-v^2}y} \\
    &- \frac{2v(1+\gamma)}{1-v^2} \zeta,\\
    -2f'&= \et \brc{\frac{2+\gamma v^2 -v^2}{1-v^2} f +
    \frac{2(1+\gamma)}{1-v^2}z},\\
    -2j'&= \et \brc{-\frac{2v(1+\gamma)}{1-v^2}f
    +\frac{1-\gamma-2v^2}{1-v^2}j - \frac{2v(1+\gamma)}{1-v^2} z}.
    \label{BC-disk}
\end{split}
\end{equation}
The equations of motion in the disk read
\begin{equation}
\begin{split}
    &2v a - \brk{v(v^2+2) +
    Q\frac{4 \gamma v^2 + \gamma+2+3v^2}{1+\gamma}} b - (v+Q) c -
    4Q vd \\
    &- 2(1+v^2+2Qv) y-\frac{2}{m} e^{P-Z/2}[2n-Qv(1-n)] z
    - 2\brk{Q\frac{1+\gamma v^2}{1+\gamma}+v} \frac{\zeta}{\et} =
    0,\\
    &(1+v^2) a - \brc{\frac{\gamma(1- v^2)}{1+\gamma}
    +4Qv+Qv^3+3v^2} b - (v+Q) vc - 2(1+v^2)Qd \\
    &- 2(Q+Qv^2+2v) y - \frac{2v}{m}e^{P-Z/2} (1+n) z - 2 \brc{Qv
    + \frac{v^2+\gamma}{1+\gamma}} \frac{\zeta}{\et}=0,\\
    &vQa+(2Q+Qv^2+2v)vb+2(Q+Qv^2+v)d +2m\frac{Q+v}{1-n} e^{Z/2-P}f\\
    &+ 2(Q+Qv^2+2v)y+2vm\frac{Q+v}{1-n}e^{Z/2-P} j + 2m\frac{Q+v}{1-n}
    e^{Z/2-P}z=0. \label{EOM}
\end{split}
\end{equation}
Equations \eqref{BC-axis}, \eqref{BC-disk} and \eqref{EOM} form
the complete set of boundary conditions.

\section{Newtonian Limit}
\label{Newtonian} As shown in \cite{Shu:2000}, the Newtonian
magnetized isothermal singular disks allow bifurcation to
non-axisymmetric equilibria if the rotation velocity is high
enough.  Specifically, the onset of bifurcation occurs at
\begin{equation}
    D^2 = \frac{m}{m+2}, \text{ for } m\ge 2,
\end{equation}
where $D$ is the ratio of rotation speed to magnetosonic speed (which
equals the sound speed at zero magnetization).
Our equation should yield the same result in the limit $v\ll 1$
and $\gamma \ll 1$. As usual, the Newtonian limit is recovered by
setting
\begin{equation*}
    g_{00} = -1 -2\Phi+ O(v^4), \quad g_{0j} = O(v^3), \qquad g_{ij} =
    \eta_{ij} + O(v^2),
\end{equation*}
where $\Phi$ is the Newtonian gravitational potential and is of
order $v^2$.  We start with the unperturbed axisymmetric solution
described by the metric \eqref{bifur-metric} and the associated
Einstein equations \eqref{unperturbed-dyn} and
\eqref{unperturbed-constraint}. The Newtonian limit becomes
\begin{equation}
\begin{split}
    &\Phi = (N/2+n\log r)(1-Q^2) = O(v^2) + O(v^4), \\
    &e^{2P-N} =
    \sin^2\theta + O(v^2), \qquad e^{P} Q = O(v^3), \qquad Z-N
    = O(v^2),
\end{split}
\end{equation}
or
\begin{equation*}
    N \sim n \sim Z \sim \Theta_0 - \frac{\pi}{2} \sim e^P -
    \sin \theta \sim \gamma \sim v^2, \text{ and } Q \sim v^3.
\end{equation*}
The equation for $N$ reads
\begin{equation*}
    N'' + N' \cot \Theta + 2n = \et \Delta, \qquad N'(0) = 0,
\end{equation*}
which has the solution
\begin{equation*}
    N' = -2n \tan \frac{\Theta}{2}, \qquad 4n \tan
    \frac{\Theta_0}{2} = \et.
\end{equation*}
Since $\Theta_0 \approx \frac{\pi}{2}$, we have $\et \approx 4n$,
which means it's also small.  Thus, the equation for $P$ becomes
\begin{equation*}
    0 = 2 \cot \Theta_0 \Rightarrow \Theta_0 \equiv \frac{\pi}{2},
    \text{ and } \Theta \equiv \theta.
\end{equation*}
$Z$ may be most directly computed through the second constraint
equation relating it to $N'$,
\begin{equation*}
    Z' = 2n N' = -4n^2 \tan \frac{\theta}{2} = O(v^4).
\end{equation*}
Finally, the last constraint equation may be solved order by
order.  The $O(v^4)$ terms are identically zero by our solution of
$N'$.  The next order is $O(v^6)$, which reads
\begin{align*}
    &(Q \cot \theta - Q')^2 - Q^2 - 2Q\cot \theta (Q \cot \theta - Q')
= 0 \\
    \Rightarrow &\log Q' = \csc \theta\\
    \Rightarrow & Q = C \tan \frac{\theta}{2}.
\end{align*}
Integrating the dynamic equation for Q, we obtain a jump
condition, which in this limit reads
\begin{equation*}
    C = 2n(C+2v) \Rightarrow C = 4nv.
\end{equation*}
Putting everything together, the limiting solution is
\begin{equation}
\begin{split}
    &n = v^2 +\gamma, \qquad \et = 4n, \qquad P = \log \sin \theta,\\
    &N' = - 2n \cot \frac{\theta}{2}, \qquad Q = 4nv \tan
\frac{\theta}{2}, \qquad  Z' = -4n^2 \tan \frac{\theta}{2}.
\end{split}
\end{equation}
Of course, in the purely Newtonian case, $Q$ and $Z$ are taken to
be $0$ since they are of higher order. A simple integration yields
\begin{equation*}
    N = -2n \log (1-\cos \theta)/2 \Rightarrow \Phi =
    -(v^2+\gamma) \log\brk{\frac{r}{2}(1-\cos \theta)},
\end{equation*}
which is the correct result for a hot Mestel disk.

In the presence of a perturbation, we still demand the Newtonian
limit to be valid.  Thus, the only nontrivial metric perturbation
is in $g_{00}$ which is $a$, and
\begin{equation*}
    g_{00} = -r^{2n} e^N (1 - a e^{im\phi}).
\end{equation*}
Expanding everything to leading order in $v$, the Einstein
equations reduce to
\begin{equation}
    -a''-\cot \theta a' + \frac{m^2}{\sin^2 \theta} a = \zeta
    \Delta.
\end{equation}
This is nothing more than the Poisson equation for the perturbed
potential $V=-a/2$.

Next, we'll derive the Newtonian version of the equations of
motion. From the Poisson equation, we know that $a$ is of order
$\zeta$, which is in turn of order $v^2$, while $y$ and $z$ are
both of order $v$. It is worthwhile to point out that we are doing
a two-parameter expansion, one in $v$ and one in the perturbation.
In this limit, the EOM becomes
\begin{equation}
\begin{split}
    &y+v\frac{\zeta}{\et} = 0,\\
    &\frac{a_0}{2} - 2vy -\frac{v}{m} z -
    (v^2+\gamma)\frac{\zeta}{\et} = 0,\\
    &2y + mz=0,
\end{split}\label{Newtonian_Bifur_EOM}
\end{equation}
where $a_0$ is evaluated on the disk, of course.  A not so trivial
calculation by \cite{Galli:2001} shows that $a_0 = \zeta/2m$.
Combining all three equations, we have (recall $\et =
4(v^2+\gamma)$)
\begin{equation*}
    \frac{v^2+\gamma}{m} + v^2 - \frac{2v^2}{m^2}-\gamma = 0
    \Rightarrow \frac{v^2}{\gamma} = \frac{m}{m+2} \text{ or }
    m=1.
\end{equation*}
This is the Newtonian bifurcation point obtained by
\cite{Shu:2000}. \footnote{The conclusion that eccentric $m=1$
bifurcations can occur at any level of disk rotation is flawed, as
are the analyses of \cite{Syer:1996, Shu:2000} and
\cite{Galli:2001} on this point (Toomre, personal communication;
Shu, Toomre, Tremaine, in preparation). Except for one special
rotation rate (the true bifurcation value), the stress tensor is
nonzero at the origin (implying a physically unrealistic steady
flow of momentum from the origin to infinity), even though it has
zero divergence.}

\section{Numerical Implementation}
\label{Numerical} Equation \eqref{Einstein} is linear in the
metric perturbations. Therefore it is ideal for finite
differencing, which transforms the differential equations into a
matrix equation. Consider a vector
\begin{equation}
    \mathbf{V} = \mathbf{a}\oplus \mathbf{b}\oplus \mathbf{c}\oplus
\mathbf{d} \oplus \mathbf{f}\oplus \mathbf{j}\oplus \mathbf{x},
\end{equation}
where
\begin{equation*}
    \mathbf{a} = (a_1, a_2, ... a_N),\qquad a_i = a(\Theta_i), \qquad
\qquad \Theta_N = \Theta_{\text{max}},
\end{equation*}
etc., and
\begin{equation*}
    \mathbf{x} = (\et_1, y, iz).
\end{equation*}
We do not include the values of the perturbation on the pole,
since they all vanish by the boundary conditions \eqref{BC-axis}.
For simplicity, let's use a uniform grid as we did in the
unperturbed solution.  Thus
\begin{equation*}
    \Theta_i = i h, \qquad h = \Theta_{\text{max}}/N.
\end{equation*}
With these definitions, the differential operators may be written
as
\begin{equation}
    a_i' = \frac{a_{i+1} - a_{i-1}}{2h}, \qquad a_i'' = \frac{a_{i+1}
-2a_i + a_{i-1}}{h^2}.
\end{equation}
It is easy to check that these expressions are accurate to second
order.  On the boundary, the situation is a little bit trickier.
To find the correct differencing scheme, let's expand the function
on the axis to second order:
\begin{align*}
    a_1 &= a_0 + h a_0' + \ahalf h^2 a_0'',\\
    a_2 &= a_0 + 2ha_0' + 2 h^2 a_0''.
\end{align*}
Taking the proper linear combinations, we have
\begin{equation}
\begin{split}
    a_0' &= \frac{4a_1-3a_0-a_2}{2h},\\
    a_0'' &=\frac{a_2-2a_1+a_0}{h^2}.
\end{split}
\end{equation}
Similarly, on the disk,
\begin{align*}
    a_{N-1} &= a_N - h a_N' + \ahalf h^2 a_N'',\\
    a_{N-2} &= a_N - 2h a_N' + 2h^2 a_N''.
\end{align*}
Thus,
\begin{equation}
\begin{split}
    a_N' &=\frac{3a_N -4a_{N-1} + a_{N-2}}{2h}, \\
    a_N'' & =\frac{a_{N-2} - 2a_{N-1}+a_N}{h^2}.
\end{split}
\end{equation}
The set of equations \eqref{Einstein} may now be written as a
matrix equation
\begin{equation}
    \mathbf{M}\cdot V = 0.\label{matrix}
\end{equation}
It is not too hard to convince oneself that the matrix
$\mathbf{M}$ is $(6N+3)\times (6N+3)$.  After finite differencing,
each component of $l_{ij}$ is represented by a $(6N+3)\times N$
submatrix, where the left-hand side is evaluated at $\Theta = h,
2h, ..(N-1)h$.  The last row in this submatrix is replaced by the
boundary condition \eqref{BC-disk}.  We thus fill the first $6N$
rows of $\mathbf{M}$. The very last three rows are the remaining
boundary conditions of \eqref{BC-axis}.

In order for \eqref{matrix} to have nontrivial solutions $V$, the
determinant of $M$ must vanish.  Schematically, the elements of
$M$ are nonlinear functions of the azimuthal quantum number $m$,
and unperturbed metric coefficients (which are, in turn, functions
of $v$ and $\gamma$).  Thus, for a given value of $m$ and
$\gamma$, the problem of stability analysis is reduced to root
finding of the equation
\begin{equation}
    |M_{m, \gamma}(v)| = 0 \label{bifur}.
\end{equation}

\section{Results}
\label{result} We scan the entire solution space looking for the
solution to \eqref{bifur}.  As expected, the bifurcation points
form two sets of tracks in the $\gamma-v^2$ space. The behaviors
of these two tracks are fundamentally different, since they are
caused by two different mechanisms.  We will discuss them
separately.

\subsection{Radiation Driven Neutral Modes}
The first set of tracks is shown in Fig \ref{rmode}.  These modes
are believed to be the analog of Rossby modes first discovered by
Chandrasekhar in 1970 and subsequently studied extensively in the context of
neutron stars. Even though our self-similar disk geometry is in
some sense infinitely different from the finite spherical geometry
of neutron stars, the underlying mechanism for these neutral modes
can still be understood if one is comfortable with the idea of
gravitational radiation with infinite wavelength. To make a bad
situation even worse, since our disks are formally infinite in
size, one is never able to reach the radiation zone to study the
gravitational wave.  However, since the original argument of
Chandrasekhar did not rely crucially on the asymptotic flatness of
spacetime, the \textit{qualitative} result still holds in this
pathological case. Consider a non-axisymmetric disturbance in the
disk which moves at a velocity $v_1<v$.  As a result, the total
angular momentum (or more appropriately, the specific angular
momentum) decreases. These perturbations will in general radiate
due to non-axisymmetry.  If in the LNRO $v_1<0$, then
gravitational radiation carries away negative angular momentum,
which damps the amplitude of perturbation, and these modes are
stable.  On the other hand, if $v_1>0$, gravitational radiation
carries away positive angular momentum, and thus the amplitude of
perturbation has to grow to make the total angular momentum more
negative.

For a given equation of state specified by $\gamma$, the
radiation-driven neutral modes occur at lower Mach number for
increasing $m$.  This is expected from the analysis of
\cite{FS:1975, FS:1978a, FS:1978b}. In fact, the onset of
instability for $m\rightarrow \infty$ occurs at zero rotational
velocity. However, for a realistic system, the strength of a
particular unstable mode is intimately related to the magnitude of
the imaginary part of its frequency.  For the $m\rightarrow
\infty$ mode, even though it is formally unstable at zero
rotation, the characteristic growth time scale is infinite. This
is due to the fact that multipole radiation is exceedingly weak
for higher values of $m$. If we truncate the self-similar disk,
the strongest unstable modes -- that is, the modes with shortest
growth time -- are still the ones with small $m$.  In the absence
of viscosity, these modes will grow until the non-linear effects
set in and limit the final rotation speed of the full disk.

In addition to the R-mode tracks, the $Q=1$ curve is also plotted
in Fig \ref{rmode}.  We would like to remind the readers that
these tracks represent models where the characteristic frequency
of a given mode becomes purely real, i.e., modes that are
marginally stable. It is not too difficult to convince oneself
that \textit{every} mode in the background of an ergoregion is
unstable. Using the simplistic picture of retrograde disturbances,
we see that in the ergoregion, every mode has to propagate in the
direction of the underlying disk as seen by an LNRO, and
gravitational radiation will drive them unstable. Furthermore, as
\cite{Friedman:1978} demonstrated, a spacetime with an ergoregion
is unstable even under scalar and vector perturbations.  From Fig
\ref{rmode}, the $m=2$ track is entirely above the $Q=1$ curve and
so is part of the $m=3$ track. In the presence of perturbing
matter fields other than those in the disk itself, the ergoregion
is likely to put a more stringent limit on the maximum rotation
rate for a disk.

The question arises whether the growth of R-modes might be
suppressed by viscous torques due to, e.g., a magneto-rotational
instability (MRI) acting in the disk \cite{MRI}. Since the viscous
effects must act to erase non-axial symmetries of length scale
$\sim r/m$, the damping time associated with linear global R-modes
must be the diffusion time scale $t_D \sim r^2/(m^2\nu)$, where
$\nu$ is the effective MRI kinematic viscosity. If relativistic
disks are even weakly magnetized and electrically conducting, as
their Newtonian counterparts are believed to be, for $m\sim 1$,
the time scale $t_D$ may be only two orders of magnitude longer
than the dynamical timescale $r/v$. R-mode spin-down for such
disks may then be effectively suppressed by the MRI viscosity, but
calculations of non-self-similar relativistic disks are needed to
answer definitively whether the growth rate of R-modes (here zero)
can overcome the viscous damping.

\subsection{Newtonian Bifurcation Track}
The second set is the generalization of Newtonian bifurcation
computed by \cite{Shu:2000}.  In Fig \ref{Newtonian_bifur}, we
plotted these extended ``Newtonian'' tracks for $m = 2,3,4,5$, and
$\infty$. The finite-$m$ values are the numerical result from
solving \eqref{bifur}. When $m$ becomes large, we may approximate
it by a continuous variable and use it as a parameter in the
asymptotic expansion. Assume all the perturbation amplitudes
remain infinitesimal in the large-$m$ limit, then the coefficients
of each power of $m$ need to vanish independently.  This
requirement translates to
\begin{equation}
\begin{split}
    &a + 2Q^2 b + Q^2 c-2d = 0, \quad Q(1+2n)f = (n+nQ^2 - Q^2) j,\\
    &-a +(1-Q^2)b + c + 2Q d = 0, \quad Qf = [Q^2(1-n)+n+2]j,\\
    &c = 0, \quad nQf = -(Q^2-1-nQ^2) j,\\
    &Qb + Qc = 0, \quad nf = Q(n-2)j,\\
    &\ahalf Q(1-n) a + Q(1-n)b - \ahalf (n+3) Qc - (1-n)d = 0, \quad
Qj = f, \quad Qf = Q^2j,\\
    &(1-n)(1+\ahalf Q^2) a + 2(1-n) Q^2 b- [1+n+\ahalf Q^2(1-n)]c -
3(1-n)Qd.
\end{split}
\end{equation}

Regardless of whether the disk is rotating, these (linear,
homogeneous) equations only have trivial solutions.  This is the
familiar Cowling approximation where the metric perturbations
vanish in the large-$m$ limit. These Cowling modes may be
understood in the following schematic way.  Recall that in the
Newtonian limit, we can invert Poisson's equation via a Green's
function, and obtain an integral representation of the
gravitational field. This procedure can also be done for the full
Einstein equations in principle, although not analytically.
Mathematically, the integral representation of metric
perturbations is effectively an average of the matter perturbation
times the Green's function. Therefore, in the $m\rightarrow
\infty$ limit, the gravitational field is indifferent to the
matter perturbation, since it averages to the axisymmetric
equilibrium over any finite angular integration. This line of
argument may be made mathematically rigorous with some more
thought, but it is not necessary here.

In the absence of a metric perturbation, the equations of motion
now simplify to
\begin{equation*}
    (1+v^2+2Qv)y + \brk{Q\frac{1+\gamma v^2}{1+\gamma} + v}
    \frac{\zeta}{\et} = 0, \quad
    (Q+Qv^2 + 2v) y + \brk{Qv + \frac{v^2+\gamma}{1+\gamma}}
    \frac{\zeta}{\et} = 0.
\end{equation*}
Eliminating the factor $\zeta/\et y$, we can combine these
equations to give
\begin{equation*}
    (1+v^2+2Qv)(Qv+Qv\gamma + v^2+\gamma)
    = (Q+Q\gamma v^2 + v+v\gamma)(Q+Qv^2+2v).
\end{equation*}
The last equation implicitly defines a surface of neutral modes
$Q(\theta_0)=f(\gamma, v)$. Recall that the solution space of
axisymmetric disks can also be viewed as a two-dimensional surface
defined by $Q(\theta_0) = g(\gamma, v)$. Thus the intersection of
these two surfaces gives the bifurcation curve in the $\gamma-v$
space.  This curve is also plotted in Fig. \ref{Newtonian_bifur}.

Near the origin, these bifurcations tracks recover the Newtonian
result, where the bifurcation point is located at
\begin{equation*}
    \frac{v^2}{\gamma} = \frac{m}{m+2}, \quad v, \gamma
    \rightarrow 0.
\end{equation*}
As $\gamma$ increases, the relativistic effects become important.
Intriguingly, the point of bifurcation occurs at lower Mach number
for a relativistic disk than in its Newtonian counterpart.  The
curve defined by $Q=1$ on the disk is again plotted. As seen in
Fig \ref{Newtonian_bifur}, these tracks are confined in the
portion of solution space where the time-like Killing vector
remains time-like. In fact, the only place where a bifurcating
neutral mode is allowed in the ergoregion is for $m=\infty$ and
$\gamma = 1$.  Even then, the spacetime is only marginally
ergo-like in the sense that $Q\rightarrow 1^-$ on the disk.  To
understand this phenomenon, we examine the velocity perturbation
corresponding to the bifurcating neutral modes.  In the Newtonian
limit, we can evaluate it analytically [see equation
\eqref{Newtonian_Bifur_EOM}, or equation (21) of \cite{Shu:2000}]
\begin{equation*}
    \delta v = y e^{im\phi} = - \frac{v}{4(v^2+\gamma)}
    \delta \et.
\end{equation*}
In general, this component of the eigenvector needs to be computed
numerically.  Actually, the exact form, or even the magnitude of
this velocity perturbation, is not important.  It suffices to know
that $\delta v$ for this mode always has the opposite sign of
$\delta \et$.  In the linear perturbation regime, where we can
still apply the superposition principle, this observation has the
following physical picture.  The full disk solution has two
components.  One is the axisymmetric equilibrium rotating at $v$
in the $+\hat \phi$ direction with energy density given by
$\varepsilon \propto \et \delta(\theta-\theta_0) /r^2$.  The
second component is the non-axisymmetric perturbation, which is a
disk of infinitesimal energy density $\delta \et$, and
infinitesimal velocity field $\delta u^\mu$.  Along the
bifurcation track, this perturbation disk is always
\textit{counter-rotating}.  The empirical result of Fig
\ref{Newtonian_bifur} leads us to postulate that the existence of
counter-rotating non-axisymmetric disturbance is another necessary
condition for bifurcation, at least for the disk geometry.  A
corollary is that the full non-axisymmetric disk spacetime can not
have a stable ergoregion. This conjecture leads naturally to the
confinement of bifurcation tracks in the portion of the solution
space without ergoregions. For models with $Q>1$ on the disk, the
time-like Killing vector becomes space-like, and thus no
counter-rotating trajectory is allowed.

Can we elevate this conjecture to apply to a more generic
relativistic non-axisymmetric equilibrium?  The answer is a
cautious yes.  Without digressing too much into the mathematical
structure of Riemannian geometry, we would like to offer the
following plausibility argument. Suppose we are able to construct
a fully nonlinear, non-axisymmetric stationary solution to the
Einstein field equations.  It can not have an event horizon, since
black holes can not have ``hair'' (in this case, ``hair'' refers
to mass multipole moments). Furthermore, in the absence of a
space-like Killing vector $\boldsymbol{\partial}_\phi$, the
non-vanishing component of $g_{tj}$ will generate a time-dependent
quadrupole moment as seen by an inertial observer.  As a result,
gravitational radiation will continue to carry away angular
momentum and energy until the system is either axisymmetric or
static.  In this aspect, the relativistic non-axisymmetric
equilibria are analogous to the Dedekind ellipsoids, where the
figure axes are static in an inertial frame, and the configuration
is supported by pressure and internal motion.  It can be shown
\citep[see, e.g.,][]{Chandra:1983} that a static metric can always
be brought to the diagonal form after appropriate coordinate
transformations.  With a diagonal metric, the ergoregion defined
by $g_{tt} = 0$ coincides with the event horizon.  Therefore, the
absence of an event horizon means a non-axisymmetric static
solution does not have an ergoregion.

\section{Summary and Discussion}

We performed a linear stability analysis of the relativistic
self-similar disk against non-axisymmetric perturbations.  For
simplicity, we restricted the class of perturbation under
consideration to be self-similar and polar.  Mathematically, this
means the scaling law is preserved and the metric is symmetric
about the midplane even in the presence of perturbation.

As expected, the Newtonian bifurcations found by \cite{Shu:2000}
and \cite{Galli:2001} have extensions into the fully relativistic
regime.  These tracks seem to exist only in models which do not
admit an ergoregion.  The corresponding velocity perturbation is
strictly negative for any positive energy density increase.  We
thus hypothesize that in addition to the existence of neutral
modes, retrograde disturbance may also be a necessary condition
for bifurcation to non-axisymmetric disk equilibria. This line of
arguments leads us to speculate that the non-axisymmetric
equilibrium solutions in general can not have ergoregions. We have
no proof that the behavior is generic, whereas the bifurcation of
rapidly rotating axisymmetric equilibria to non-axisymmetric forms
probably is.

In addition, we have discovered the onset of R-mode instability,
which is driven by gravitational radiation. The marginal stability
tracks follow the qualitative behavior first discussed by
\cite{FS:1978b}, and the result here is probably also generic. For
a self-similar disk, the entire $m=2$ neutral-mode tracks and part
of $m=3$ occur in models with ergocones.  We believe that in
general for a fixed equation of state, the onset of instability
occurs either at the ergoregion formation, or the tracks we
computed, whichever have lower velocity.

\begin{figure}[ht]
\begin{center}
\includegraphics[height = .8\textwidth, angle = 90]{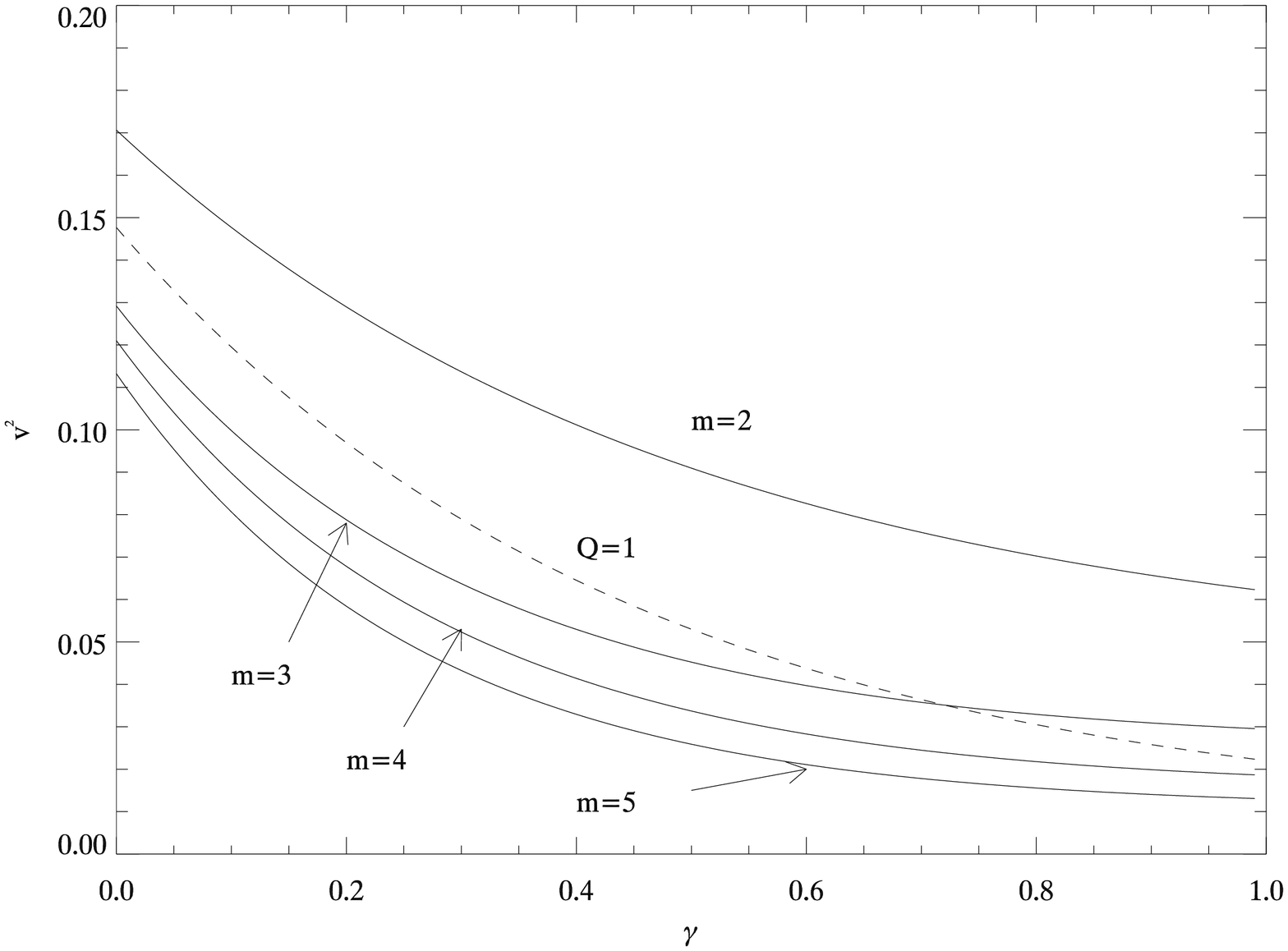}
\caption{Radiation driven neutral modes. For large enough values
of $m$, the onset of instability is believed to occur at
infinitesimal velocities.}\label{rmode}
\end{center}
\end{figure}

\begin{figure}[ht]
\begin{center}
\includegraphics[height = .8\textwidth, angle = 90]{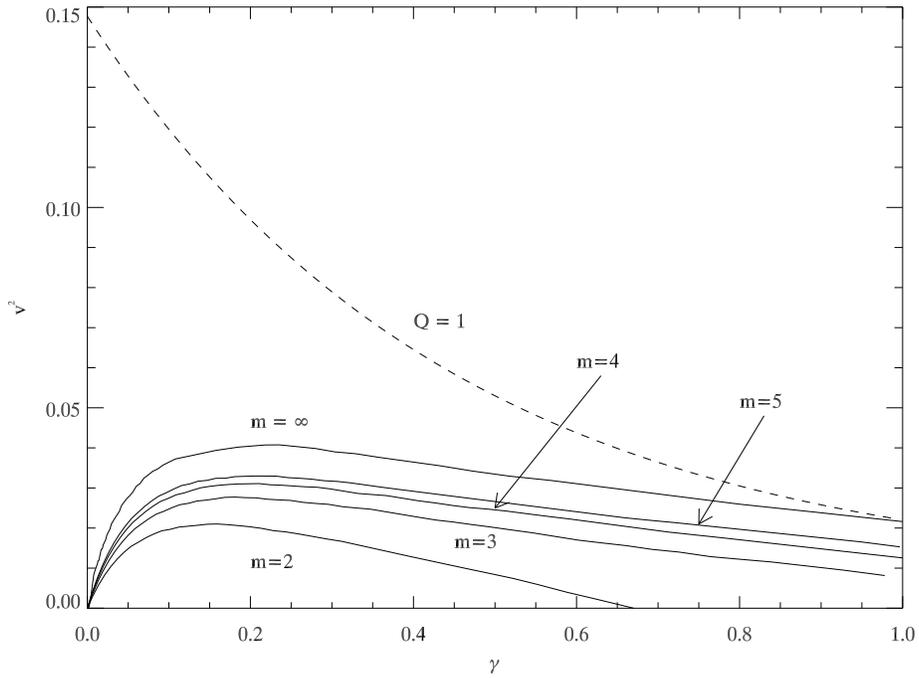}
\caption{The neutral mode curves which are connected to the
Newtonian bifurcation.  In the lower-left corner, where
$\gamma\rightarrow 0$ and $v \rightarrow 0$, the slope is
$m/(m+2)$, as expected from the Newtonian limit.  The dashed line
separates the solution space into ergoregion and non-ergoregion.}
\label{Newtonian_bifur}
\end{center}
\end{figure}

These studies are the first step to constructing fully
non-axisymmetric relativistic equilibria.  For a given value of
$\gamma$, if the axisymmetric state contracts quasi-statically by
shedding angular momentum, the linear rotational velocity will
increase.  This evolution represents a vertical line in Figs
\ref{rmode} and \ref{Newtonian_bifur}. Eventually, the velocity
will reach a value where a non-axisymmetric mode becomes unstable.
If they undergo gravitational collapse, the secular instability
will most likely survive over many dynamic time scales (which, in
the purely self-similar case, is infinite). Therefore, the
collapse will be fundamentally non-axisymmetric.  For simplicity,
we have only considered each Fourier component independently. In
the linear perturbation regime, the effect of a general
non-axisymmetric disturbance can always be decomposed into its
Fourier components with each component decoupled from others. When
the amplitudes become finite, our Fourier series will fail to
converge, and a more detailed analysis is required.

The current work serves as a spring board to one of the ultimate
challenges in numerical relativity -- the fully nonlinear
numerical simulation of a non-axisymmetric collapse. Only then can
we answer questions such as whether the central singularities of
objects like relativistic SIDs remain naked.

We thank the referee for valuable comments.  This work has been
supported by Academia Sinica through the Distinguished
Postdoctoral Research Fellowship awarded to MJC, and by National
Science Council through grants NSC 91-2112-M-007-29 and NSC
92-2112-M-001-062 awarded to FHS.

\newpage

\end{document}